\documentclass[useAMS,usenatbib]{mn2e}
\usepackage[utf8]{inputenc}
\usepackage[T1]{fontenc}
\usepackage{epsfig}
\usepackage{graphicx}
\usepackage{graphics}
\usepackage{color}
\usepackage{amssymb}
\usepackage{amsmath}
\usepackage{subfig}
\usepackage{hyperref}
\usepackage{caption}

\newcommand{\aap}{Astronomy and Astrophysics}

\newcommand{\mnras}{Monthly Notices of the Royal Astronomical Society}
\newcommand{\apj}{The Astrophysical Journal}
\newcommand{\apjs}{Astrophysical Journal Supplements}
\newcommand{\apjl}{Astrophysical Journal Letters}
\newcommand{\aj}{Astronomical Journal}
\newcommand{\nat}{Nature}
\newcommand{\araa}{Annual Review of Astronomy and Astrophysics}

\newcommand{\na}{New Astronomy}

\newcommand{\rmxaa}{Revista Mexicana de Astronomía y Astrofísica}
\newcommand{\sovast}{Soviet Astronomy}
\newcommand{\pasa}{Publications of the Astronomical Society of Australia}

\newcommand{\pra}{Physical Review A (Atomic, Molecular, and Optical Physics)}

\title[Supersoft Source Nebulae]
  {Where are all of the nebulae ionized by supersoft X-ray sources?}
\author[T.E. Woods and M. Gilfanov]{T.E. Woods$^{1,2}$\thanks{E-mail:
tewoods.astro@gmail.com} and M. Gilfanov$^{1,3}$\\ $^{1}$Max Planck Institute for Astrophysics, Karl-Schwarzschild-Str.
1, Garching b. M{\" u}nchen 85741, Germany\\
$^{2}$Monash Centre for Astrophysics, School of Physics and Astronomy, Monash University, VIC 3800, Australia\\
$^{3}$Space Research Institute of Russian Academy of Sciences, Profsoyuznaya 84/32,117997 Moscow, Russia\\}
\pagerange{\pageref{firstpage}--\pageref{lastpage}}

\def\LaTeX{L\kern-.36em\raise.3ex\hbox{a}\kern-.15em
    T\kern-.1667em\lower.7ex\hbox{E}\kern-.125emX}

\hypersetup{
   pdfauthor={Tyrone E Woods, Marat Gilfanov},
   pdftitle ={Supersoft Source Nebulae}
}

\begin{document}

\label{firstpage}

\maketitle

\begin{abstract}
Accreting, steadily nuclear-burning white dwarfs are associated with so-called close-binary supersoft X-ray sources (SSSs), observed to have  temperatures of a few$\times 10^{5}$K and luminosities on the order of $10^{38}$erg/s. These and other types of SSSs are expected to be capable of ionizing their surrounding circumstellar medium; however, to date only one such nebula was detected in the Large Magellanic Cloud (of its 6 known close-binary SSSs), surrounding the accreting, nuclear-burning WD CAL 83. This has led to the conclusion that most SSSs cannot have been both luminous ($\gtrsim 10^{37}$erg/s) and hot ($\gtrsim$ few $\times 10^{4}$K) for the majority of their past accretion history, unless the density of the ISM surrounding most sources is much less than that inferred for the CAL 83 nebula (4--10$\rm{cm}^{-3}$). Here we demonstrate that most SSSs must lie in much lower density media than CAL 83. Past efforts to detect such nebulae have not accounted for the structure of the ISM in star-forming galaxies and, in particular, for the fact that most of the volume is occupied by low density warm \& hot ISM. CAL 83 appears to lie in a region of ISM which is at least $\sim 40$-fold overdense. We compute the probability of such an event to be $\approx 18\%$, in good agreement with observed statistics. We provide a revised model for the ``typical'' SSS nebula, and outline the requirements of a survey of the Magellanic clouds which could detect the majority of such objects. We then briefly discuss some of the possible implications, should there prove to be a large population of previously undiscovered ionizing sources.

\end{abstract}


\section{Introduction}

Supersoft X-ray sources (SSSs) comprise a unique and varied class of objects, first identified more than two decades ago by the Einstein observatory (HEAO-2). Close binary SSSs are characterized by significant emission in the 0.3--0.7 keV spectral band resembling the Wien tail of a blackbody, with $\rm{T}_{\rm{eff}}$ $\approx$ 2--10 $\cdot 10^{5}K$ and estimated $\rm{L}_{\rm{bol}}$ $\sim$ $10^{37-38}$ erg/s \citep{Greiner2000}. If such high temperatures and luminosities are typical of the majority of their accretion lifetimes ($\sim 10^{6}$yrs), then these objects should strongly ionize any surrounding interstellar medium (ISM), producing characteristic emission-line nebulae \citep{Rappaport94}. However, to date only one such nebula (CAL 83) has been identified \citep{PM89,Remillard95}. In the absence of a compelling reason for all other SSSs to be surrounded by a more rarefied ISM than CAL 83, the lack of high-ionization nebulae accompanying other known sources would imply an extremely low duty-cycle for these objects \citep{CR96}. At present there is little understanding as to how either possibility would come about. This uncertainty has remained unresolved over the last two decades.

Several varieties of accreting compact objects are thought to undergo a super-soft or quasi-soft X-ray emitting state, including accreting black holes and post-outburst novae. Among the most common objects in this class, at least as found in the Magellanic clouds \citep{Greiner2000}, are the so-called close-binary SSSs. In the classic picture \citep{vdHeuvel92}, close-binary SSSs are white dwarfs (WDs) accreting hydrogen-rich matter from a main sequence or red giant companion, with steady nuclear burning of the accreted hydrogen at the WD's surface. This steady nuclear burning is only possible for mass transfer rates within a narrow range \citep[1 -- few $\cdot$ $10^{-7}M_{\odot}/$yr, e.g.][]{Nomoto07}. For accretion rates less than this, nuclear-burning proceeds only in unsteady bursts, giving rise to novae. The upper bound is approximately given by the accretion rate at which a WD has the same nuclear-burning luminosity as an AGB star with the equivalent core mass. Above this threshold, the observational appearance is extremely uncertain. One possibility is that the photosphere of the WD will once more swell up to the thermal equilibrium radius of a giant \citep[see e.g.][]{Cassisi98}. However, dynamical processes, such as an optically-thick wind driven from within the photosphere, may modulate mass transfer before this can happen, with photospheric temperatures in the range 1--2 $\cdot$ $10^{5}K$ \citep{HKN99} for $\dot M \lesssim 3\times 10^{-6}\rm{M}_{\odot}$/yr. 

Apart from simply understanding their phenomenology, close-binary SSSs have come under particular scrutiny as candidates for the progenitors of thermonuclear (type Ia) supernovae (SNe Ia). In the so-called single degenerate channel \citep{WI73}, a carbon-oxygen WD accretes hydrogen-rich material as described above until reaching the Chandrasekhar mass ($\rm{M}_{\rm{Ch}} \approx 1.38\rm{M}_{\odot}$), triggering a thermonuclear runaway \citep[although both lower and higher explosion masses may be possible, see][]{MMN14}.

A considerable body of evidence has accumulated in the past several years demonstrating that such sources cannot be the progenitors of SNe Ia, or, should the SD scenario still hold, that accreting WD progenitors cannot manifest themselves as SSSs for the majority of their accretion phase \citep{GB10, DiStefano10}. Such arguments were predicated on the notion that steadily nuclear-burning sources with radii typical of massive WDs should emit strongly in the soft X-ray band ($\rm{T}_{\rm{eff}}\approx 5\times 10^{5}$--$10^{6}$K), producing detectable emission where the intervening column density is low (e.g. along many lines of sight in the Magellanic Clouds, M31, and nearby early-type galaxies). 

However, for lower WD masses \citep[M $\lesssim 1\rm{M}_{\odot}$,][]{WB13}, or in case of  considerable inflation of the photosphere above that expected from the WD mass -- radius relation \citep{Panei00}, the temperature may fall below $\sim 5\times 10^{5}$K. This would shift the majority of the nuclear-burning luminosity to the extreme-UV, where it could not be directly detected. However, such objects should remain very strong sources of ionizing radiation \citep{Rappaport94}, with no cut-off at the H I or He II ionization edges \citep{Rauch10}. One can then demonstrate that, in the single degenerate scenario, accreting nuclear-burning WDs must produce an enormous combined ionizing luminosity, easily sufficient to dominate the ionizing background in many early-type galaxies \citep{WG13, WG14}. Given that many early-type galaxies possess extended emission-line regions powered by the continuum from their old stellar populations, we should then be able to detect emission lines characteristic of ionization by high-temperature sources, such as He II 4686\AA\ and [O I] 6300\AA. Following on this, \cite{Johansson14} produced carefully-selected stacks of early-type galaxies with emission-lines in the SDSS, and found a He II 4686\AA/H$\beta$ ratio ($\approx$8\%) consistent with $\lesssim 10$\% contribution from any high-temperature ($\approx 1.2$--$6\times 10^{5}$K) sources for galaxies with luminosity-weighted mean stellar ages in the range 1Gyr $<$ t $<$ 4Gyr. Similar constraints from the [O I] 6300\AA\ line extend this limit up to T $\approx 10^{6}$K SD progenitors (Johansson et al., in prep.). Note that at these high temperatures (exceeding $\sim 5 \times 10^{5}K$), the contribution of accreting WDs is also very tightly constrained in soft X-rays \citep[$\lesssim$1\%][]{GB10}. 

In order to remain consistent with the observational constraints discussed above, the average mass accreted per WD in any ``hot mode'' ($\rm{T}_{\rm{eff}}$ $\gtrsim 1.5 \times 10^{5}$K) by SD progenitors is limited to $10^{-2}$--$10^{-3}\rm{M}_{\odot}$. However, accreting WDs may remain viable as the progenitors of at least some SNe Ia if they accrete principally at lower temperatures. Indeed, known SSSs are understood to undergo considerable variability, with many interchanging between optical-low/X-ray-on (hot) and optical-high/X-ray-off (cold) states. Searches for nebulae surrounding SSSs in the Magellanic clouds have detected only one, CAL 83 \citep{Remillard95}, which itself has proven difficult to model \citep{Gruyters12}. This implies that either the density of the ISM surrounding other SSSs is much less than that in the vicinity of CAL 83 \citep[n $\sim$ 4 -- 10$\rm{cm}^{-3}$,][]{Remillard95}, or that these sources do not radiate at high temperatures for the majority of their accretion history. For close-binary SSSs, this would be in keeping with the possibility that accreting, nuclear-burning WDs may still be a strong candidate for the progenitors of SNe Ia.

Here, we briefly review the physics of ionized nebulae surrounding high-temperature and variable sources in $\S$\ref{SSS_Stromgren}, and provide detailed predictions for the observable characteristics of such nebulae in very low-density media. We then discuss the meaning of the apparent absence of SSS nebulae in the Magellanic clouds in $\S$\ref{SSSs_MCs}. We revisit the upper limits placed by \cite{Remillard95}, which do not constrain the temperature/luminosity of SSSs in low-density media ($\rm{n}_{\rm{ISM}}$ $\sim$ 0.1 -- 1$\rm{cm}^{-3}$). We demonstrate however that such densities are typical of the warm-neutral/warm-ionized ISM, in which most SSSs are expected to be located. Based on the current understanding of the structure of the interstellar medium in star-forming galaxies, we compute the probability of finding a source in the vicinity of a cold and neutral cloud big enough to be detected by the \cite{Remillard95} survey in the Large Magellanic Cloud. We then outline the requirements for a survey which can provide a more complete census of SSS nebulae in the Magellanic clouds in $\S$\ref{Prospects}, and place this in the context of the only observed SSS nebula, as well as two candidate nebulae in M33. Before closing, we briefly discuss the plausibility and implications of a large, heretofore undiscovered ionizing source population for stellar feedback and the evolution of the ISM in $\S$\ref{Implications}.

\section{Str{\" o}mgren regions surrounding accreting WDs}\label{SSS_Stromgren}

\subsection{The classical picture}
 
\begin{figure}
\begin{center}
\includegraphics[height=0.25\textheight]{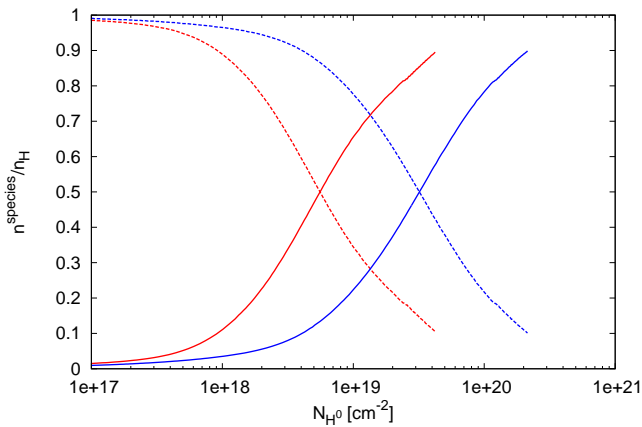}
\caption[Ionization structure (single source)]{Ionized (dotted line) and neutral (solid line) fractions of the total hydrogen density, as a function of the total column density in neutral hydrogen (integrated outward from the central source). Blue and red lines denote constant ISM densities of 0.1 and 1 $\rm{cm}^{-3}$ respectively. In both cases, we assume ionization by a central blackbody source with temperature 5$\times 10^{5}$K and bolometric luminosity $10^{37.5}$erg/s. Note that we terminate our calculations when the temperature of the gas falls below 3000K.}
\label{ionization_structure}
\end{center}
\end{figure}

The notion of emission-line nebulae ionized by SSSs was first explored by \cite{Rappaport94}. They identified several tell-tale signatures which would allow for the identification of such objects, namely pronounced [O III] 5007\AA, [O I] 6300\AA, and He II 4686\AA\ emission, similar to the most luminous planetary nebulae (PNe). However, contrary to PNe, these would be very extended objects, with typical Str{\" o}mgren radii:

\begin{equation}
\rm{R}_{\rm{S}} \approx 35\rm{pc}\left(\frac{\dot N_{\rm{ph}}}{10^{48}\rm{s}^{-1}}\right)^{\frac{1}{3}}\left(\frac{\rm{n}_{\rm{ISM}}}{1\rm{cm}^{-3}}\right)^{-\frac{2}{3}}\label{RS}
\end{equation}

\noindent where $\dot N_{\rm{ph}}$ is the ionizing photon luminosity of the source, and $\rm{n}_{\rm{ISM}}$ is the density of the surrounding ISM. In practice, nebulae ionized by high-temperature sources extend to somewhat larger radii than even this, as high energy photons penetrate deeper into the neutral ISM, yielding a broad interface at the Str{\" o}mgren boundary \citep[see e.g.][also fig. \ref{ionization_structure}]{WG14}. Such nebulae may be resolved even by ground-based telescopes out to distances of a few Mpc.

Emission-line surveys represent an attractive alternative to searching for SSSs directly for a variety of reasons \citep{Rappaport94, DiStefano95}. While soft-band ($\approx$ 0.3--0.7keV) X-ray sources may be completely obscured by relatively modest column densities ($\gtrsim$ few$\times 10^{21}\rm{cm}^{-2}$), optical emission-line nebulae typically suffer only slight reddening. The most [O III] 5007\AA- and He II 4686\AA-luminous nebulae should also be those ionized by relatively low-temperature WDs ($\approx$1--5$\times10^{5}$K), probing the SSS population in the temperature regime least sensitive to soft X-ray observations \citep{DiStefano95}.

Furthermore, if the progenitors of some SNe Ia do undergo a SSS phase, any nebula ionized by the accreting WD will remain for 

\begin{equation}
\tau _{\rm{rec}} = 1/(\alpha _{\rm{B}}\rm{n}_{\rm{ISM}}) \approx 10^{5} \left(\frac{\rm{n}_{\rm{ISM}}}{1\rm{cm}^{-3}}\right)^{-1} \rm{years} \label{timescale}
\end{equation}

\noindent after the end of this phase, potentially providing a test of the progenitors of relatively young (t $\lesssim$ $\tau _{\rm{rec}}$), nearby SN Ia remnants long after their explosion \citep[i.e. without the need for pre-supernova images, as in][]{GMS14}. Outside of the shell shocked by the supernova itself, such a ``fossil nebula'' \citep[first discussed for O stars in][]{BKS70} would remain largely unperturbed by the light of the supernova, as neither the initial shock breakout nor radiation due to $^{56}\rm{Ni}$ decay will appreciably ionize the surrounding ISM \citep[see discussion in][]{Cumming96}, and the UV tail of SNe Ia spectra suffer from considerable line-blanketing, with little flux below $\approx$1000\AA\ \citep[for a comparison of NLTE atmospheres with observed spectra of supernovae, see e.g.][]{Pauldrach96}.

\subsection{Modeling simple SSS nebulae}\label{Model_SSS_Nebulae}

\begin{figure*}
\begin{center}
\begin{tabular}{cc}
\includegraphics[height=0.25\textheight]{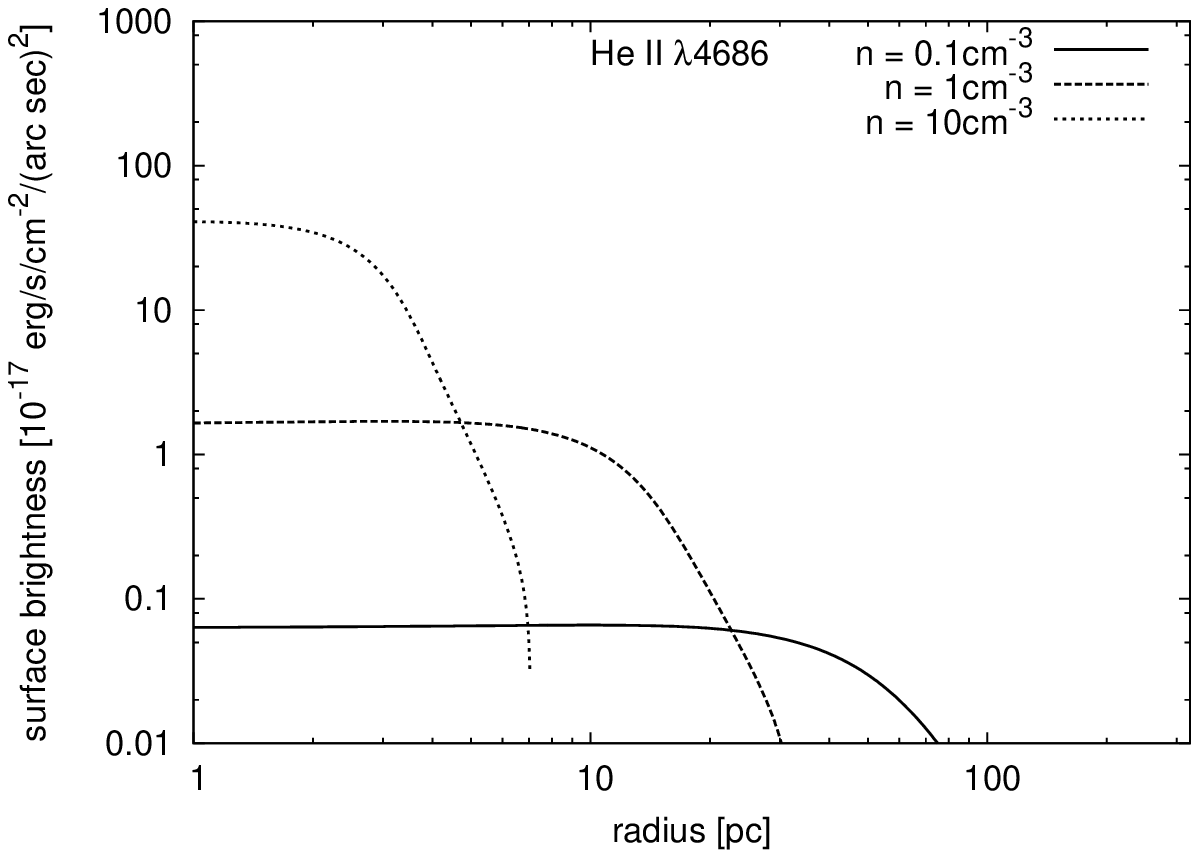} &
\includegraphics[height=0.25\textheight]{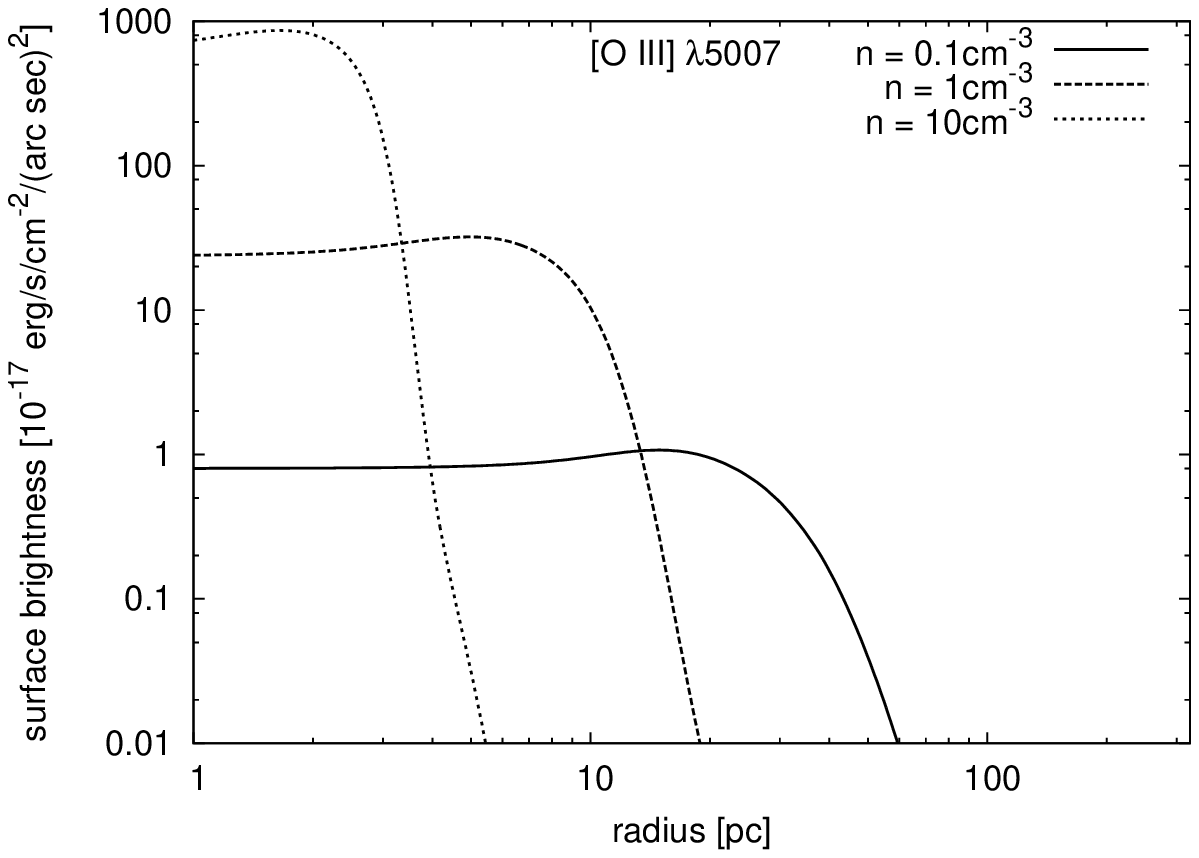}\\
\includegraphics[height=0.25\textheight]{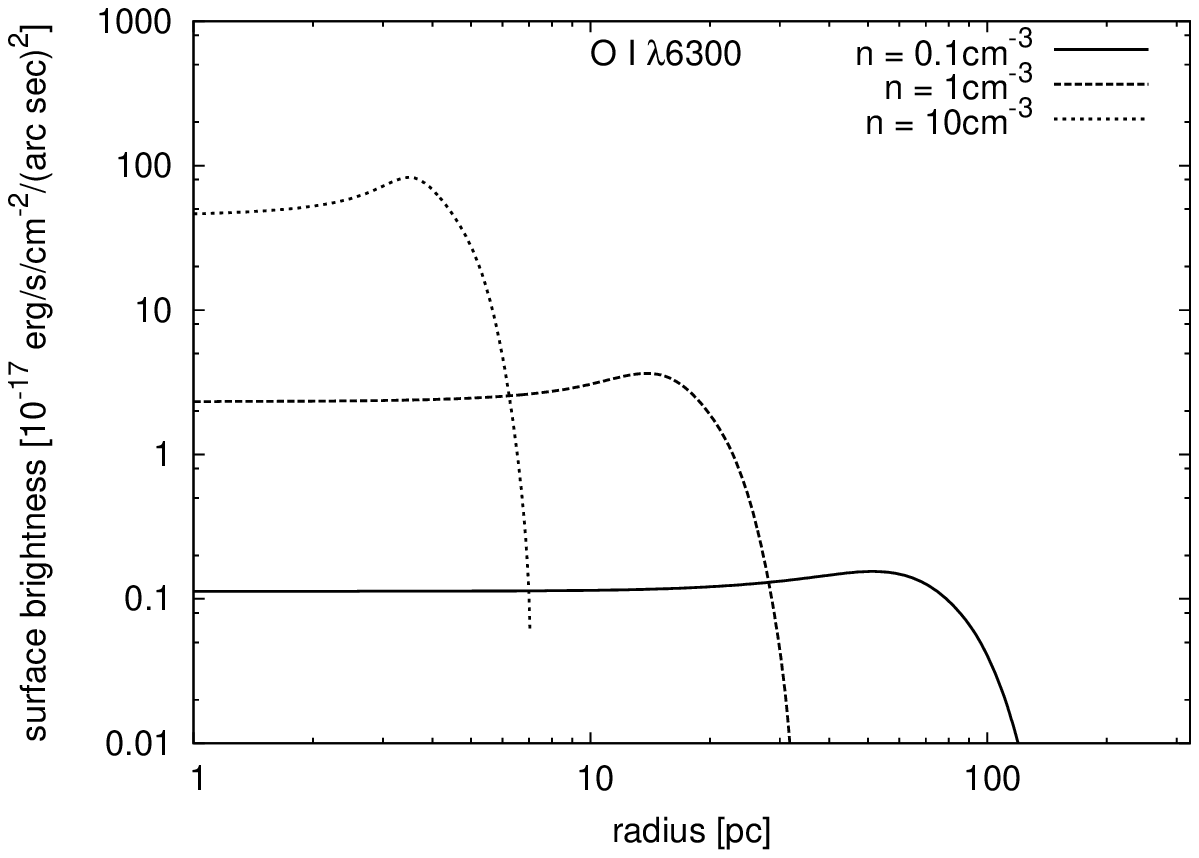}&
\includegraphics[height=0.25\textheight]{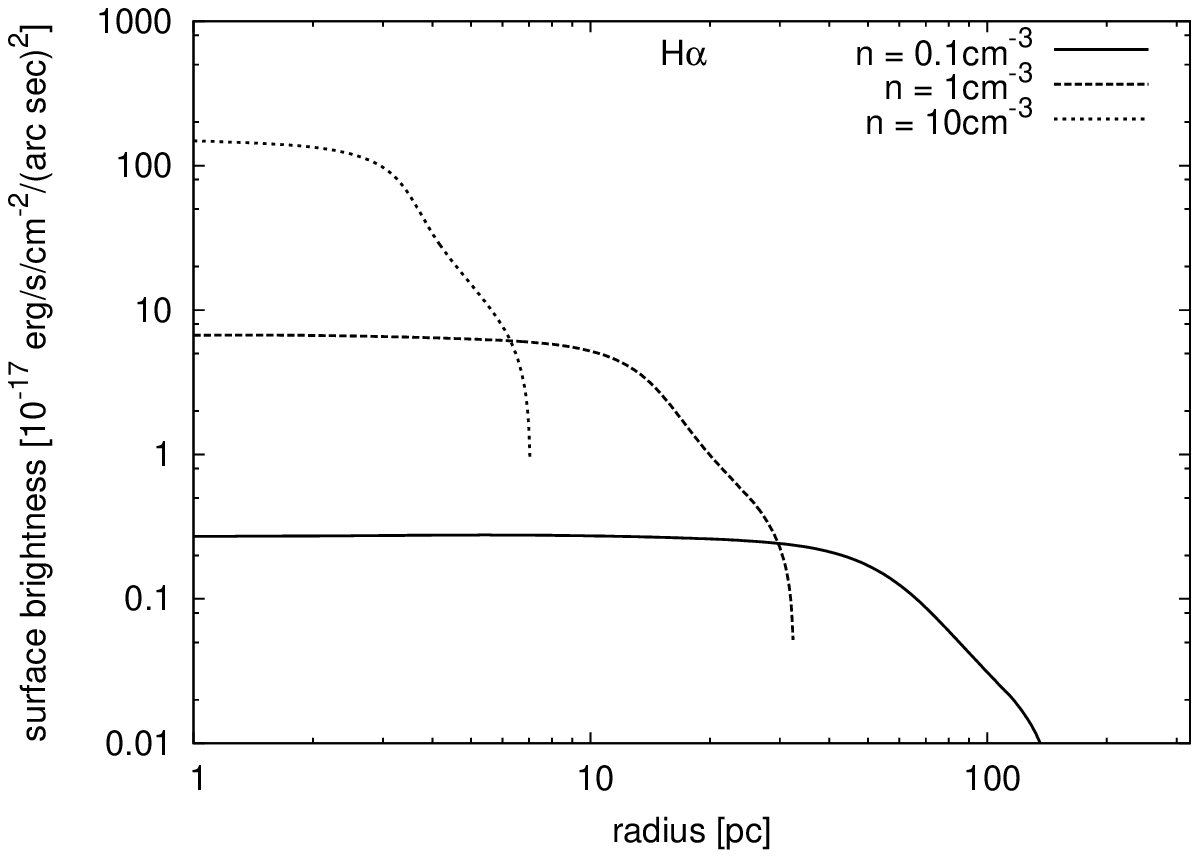}\\
\end{tabular}
\caption[Surface brightness profiles]{Surface brightness profiles for four optical emission lines assuming ionization by a central blackbody source in the LMC with temperature 5$\times 10^{5}$K and bolometric luminosity $10^{37.5}$erg/s. In each panel, the surface brightness is shown for a given emission line for differing constant ISM densities (0.1, 1, and 10 $\rm{cm}^{-3}$).
}\label{SB_profiles}
\end{center}
\end{figure*}

\begin{table*}
\renewcommand{\arraystretch}{1.1}
\begin{tabular}{c c c c c c c c c c c}
\hline
n & T & L & $\rm{R}_{H II, 0.5}$ & $\rm{R}_{He II, 0.5}$ & He II & [O III] & [O I] & H$\alpha$ & [N II] \\
$\rm{cm}^{-3}$ & $10^{5}$K & $10^{38}$erg/s & pc & pc & erg/s & erg/s & erg/s & erg/s & erg/s \\
\hline
 10 & 2 & 1    & 7.7 & 7.12 & 3.82e+35 & 7.82e+36 & 2.67e+35 & 1.98e+36 & 1.8e+36\\
 10 & 2 & 0.5  & 6.1 & 5.58 & 1.88e+35 & 3.61e+36 & 1.58e+35 & 9.88e+35 & 9.96e+35\\
 10 & 2 & 0.1  & 3.53 & 3.16 & 3.52e+34 & 5.6e+35 & 4.44e+34 & 1.94e+35 & 2.4e+35\\
 10 & 4 & 1    & 6.48 & 6.35 & 2.78e+35 & 5.24e+36 & 6.81e+35 & 1.26e+36 & 1.99e+36\\
 10 & 4 & 0.5  & 5.08 & 5.07 & 1.32e+35 & 2.29e+36 & 3.62e+35 & 6.14e+35 & 1.04e+36\\
 10 & 4 & 0.1  & 2.87 & 2.96 & 2.31e+34 & 3.06e+35 & 7.83e+34 & 1.14e+35 & 2.19e+35\\
 10 & 7 & 1    & 5.06 & 6.04 & 1.31e+35 & 2.22e+36 & 8.78e+35 & 8.22e+35 & 1.64e+36\\
 10 & 7 & 0.5  & 3.91 & 4.76 & 6.09e+34 & 9e+35 & 4.21e+35 & 3.87e+35 & 7.91e+35\\
 10 & 7 & 0.1  & 2.14 & 2.61 & 9.92e+33 & 1.02e+35 & 7.2e+34 & 6.47e+34 & 1.36e+35\\
 10 & 10 & 1   & 4.12 & 5.14 & 6.93e+34 & 1.05e+36 & 7.7e+35 & 5.89e+35 & 1.24e+36\\
 10 & 10 & 0.5 & 3.15 & 4.05 & 3.14e+34 & 4.05e+35 & 3.48e+35 & 2.66e+35 & 5.63e+35\\
 10 & 10 & 0.1 & 1.67 & 2.15 & 4.85e+33 & 4.09e+34 & 5.23e+34 & 4.07e+34 & 8.52e+34\\
 1 & 2 & 1     & 35.5 & 31.6 & 3.52e+35 & 5.54e+36 & 4.35e+35 & 1.94e+36 & 2.37e+36\\
 1 & 2 & 0.5   & 28.1 & 25 & 1.7e+35 & 2.42e+36 & 2.46e+35 & 9.61e+35 & 1.26e+36\\
 1 & 2 & 0.1   & 16.2 & 15.1 & 3.03e+34 & 3.27e+35 & 6.15e+34 & 1.86e+35 & 2.71e+35\\
 1 & 4 & 1     & 28.9 & 29 & 2.31e+35 & 3.05e+36 & 7.69e+35 & 1.13e+36 & 2.17e+36\\
 1 & 4 & 0.5   & 22.6 & 22.6 & 1.08e+35 & 1.24e+36 & 3.91e+35 & 5.45e+35 & 1.08e+36\\
 1 & 4 & 0.1   & 12.6 & 12.4 & 1.8e+34 & 1.41e+35 & 7.67e+34 & 9.77e+34 & 1.99e+35\\
 1 & 7 & 1     & 21.6 & 24.4 & 9.91e+34 & 1.01e+36 & 7.07e+35 & 6.41e+35 & 1.35e+36\\
 1 & 7 & 0.5   & 16.7 & 18.3 & 4.5e+34 & 3.85e+35 & 3.25e+35 & 2.94e+35 & 6.22e+35\\
 1 & 7 & 0.1   & 8.98 & 8.9 & 6.95e+33 & 3.75e+34 & 5.13e+34 & 4.67e+34 & 9.66e+34\\
 1 & 10 & 1    & 17 & 20.6 & 4.84e+34 & 4.08e+35 & 5.13e+35 & 4e+35 & 8.45e+35\\
 1 & 10 & 0.5  & 13 & 15.2 & 2.15e+34 & 1.48e+35 & 2.24e+35 & 1.77e+35 & 3.69e+35\\
 1 & 10 & 0.1  & 6.8 & 7.2 & 3.13e+33 & 1.32e+34 & 3.11e+34 & 2.55e+34 & 5.13e+34\\
0.5 & 2 & 1    & 56.2 & 50 & 3.39e+35 & 4.83e+36 & 4.91e+35 & 1.92e+36 & 2.51e+36\\
0.5 & 2 & 0.5  & 44.4 & 40.1 & 1.63e+35 & 2.08e+36 & 2.74e+35 & 9.5e+35 & 1.31e+36\\
0.5 & 2 & 0.1  & 25.5 & 24.1 & 2.84e+34 & 2.66e+35 & 6.55e+34 & 1.83e+35 & 2.72e+35\\
0.5 & 4 & 1    & 45.1 & 43.9 & 2.16e+35 & 2.48e+36 & 7.78e+35 & 1.09e+36 & 2.15e+36\\
0.5 & 4 & 0.5  & 35.3 & 33.6 & 1.01e+35 & 9.9e+35 & 3.91e+35 & 5.22e+35 & 1.05e+36\\
0.5 & 4 & 0.1  & 19.6 & 18.3 & 1.65e+34 & 1.07e+35 & 7.45e+34 & 9.28e+34 & 1.87e+35\\
0.5 & 7 & 1    & 33.2 & 35.1 & 8.99e+34 & 7.68e+35 & 6.48e+35 & 5.87e+35 & 1.24e+36\\
0.5 & 7 & 0.5  & 25.6 & 25.3 & 4.05e+34 & 2.86e+35 & 2.95e+35 & 2.67e+35 & 5.63e+35\\
0.5 & 7 & 0.1  & 13.6 & 12.7 & 6.15e+33 & 2.69e+34 & 4.55e+34 & 4.2e+34 & 8.48e+34\\
0.5 & 10 & 1   & 25.9 & 27.5 & 4.29e+34 & 2.96e+35 & 4.45e+35 & 3.53e+35 & 7.35e+35\\
0.5 & 10 & 0.5 & 19.7 & 19.8 & 1.89e+34 & 1.06e+35 & 1.92e+35 & 1.55e+35 & 3.18e+35\\
0.5 & 10 & 0.1 & 10.2 & 9.58 & 2.69e+33 & 9.21e+33 & 2.63e+34 & 2.2e+34 & 4.31e+34\\
0.1 & 2 & 1    & 163 & 151 & 3.03e+35 & 3.26e+36 & 6.12e+35 & 1.86e+36 & 2.7e+36\\
0.1 & 2 & 0.5  & 128 & 122 & 1.42e+35 & 1.33e+36 & 3.27e+35 & 9.18e+35 & 1.36e+36\\
0.1 & 2 & 0.1  & 72.6 & 74.4 & 2.38e+34 & 1.53e+35 & 7.03e+34 & 1.75e+35 & 2.56e+35\\
0.1 & 4 & 1    & 127 & 113 & 1.8e+35 & 1.41e+36 & 7.64e+35 & 9.76e+35 & 1.98e+36\\
0.1 & 4 & 0.5  & 98.2 & 87 & 8.26e+34 & 5.38e+35 & 3.72e+35 & 4.65e+35 & 9.35e+35\\
0.1 & 4 & 0.1  & 53.1 & 47.6 & 1.29e+34 & 5.3e+34 & 6.54e+34 & 8.04e+34 & 1.52e+35\\
0.1 & 7 & 1    & 90 & 79.4 & 6.95e+34 & 3.76e+35 & 5.1e+35 & 4.66e+35 & 9.64e+35\\
0.1 & 7 & 0.5  & 68.5 & 59.3 & 3.08e+34 & 1.35e+35 & 2.27e+35 & 2.1e+35 & 4.24e+35\\
0.1 & 7 & 0.1  & 35.3 & 31.6 & 4.45e+33 & 1.19e+34 & 3.29e+34 & 3.21e+34 & 5.94e+34\\
0.1 & 10 & 1   & 68.3 & 62.2 & 3.13e+34 & 1.33e+35 & 3.1e+35 & 2.54e+35 & 5.11e+35\\
0.1 & 10 & 0.5 & 51.3 & 45.5 & 1.35e+34 & 4.62e+34 & 1.31e+35 & 1.09e+35 & 2.16e+35\\
0.1 & 10 & 0.1 & 25.6 & 23.6 & 1.83e+33 & 3.84e+33 & 1.71e+34 & 1.53e+34 & 2.75e+34\\
\end{tabular}
\caption[Predicted luminosity of optical emission lines]{Predicted luminosities reprocessed into the strongest optical emission lines, as well as the 4$\rightarrow$3 transition of He II at 4686\AA, for sources with varying temperatures and luminosities, and embedded in surrounding ISM of varying density. Here [O III] is the 5007\AA\ line, [O I] denotes the 6300\AA\ feature, and [N II] is the combined luminosity of the 6548\AA\ and 6584\AA\ lines. Also shown are the radii (measured from the central source) at which the Hydrogen ionization fraction is 50\%, and at which 50\% of Helium is in the singly-ionized state. \label{lines}}
\end{table*}

In order to understand the observability of SSS nebulae, we must make an estimate of their typical surface brightnesses, which will depend not only on their time-averaged temperatures and luminosities, but also on the density of their surrounding ISM. In the following, we model nebulae ionized by high-temperature, luminous SSSs using the photoionization code Cloudy \citep[v13.03,][]{Cloudy_ref13}. For simplicity and ease of comparison with the results of \cite{Rappaport94}, we assume isochoric (constant density) ISM, solar gas phase abundances, and blackbody ionizing spectra throughout. The latter is well-justified, at least near the H I and He II ionization edges, by comparison with detailed NLTE calculations of SSS spectra \citep[e.g.][]{Rauch10}. At the outer boundary of our nebular models, we terminate our calculations when the gas temperature falls below 3000K (seen as the abrupt cutoff at an ionization fraction of $\approx$10\% in fig. \ref{ionization_structure}). This is easily sufficient to compute the total luminosity reprocessed in the most important diagnostic lines (particularly [O III] 5007\AA, see fig. \ref{SB_profiles}), and allows us to avoid making any assumption about the incident cosmic ray spectrum (which would become important in this part of the nebula).

This allows us to compute the volume emissivity $\epsilon(\rm{r})$ (assuming spherical symmetry) in any emission line as a function of radius, which we may then use to find the surface brightness in line i:

\begin{equation}
\rm{SB}_{\rm{i}}(r) = \int_{l} \frac{\epsilon_{\rm{i}}(r)}{4\pi} dl
\end{equation}

\noindent where we have integrated along the line of sight l.

Shown in figure \ref{SB_profiles} are the surface brightness profiles in four lines (He II 4686\AA, [O III] 5007\AA, [O I] 6300\AA, and H$\alpha$) for a nebula ionized by a SSS in the LMC as a function of radius, for varying ISM densities (0.1, 1, and 10$\rm{cm}^{-3}$) and fixed source temperature and luminosity ($5\times10^{5}$K and $10^{37.5}$erg/s). The most prominent signature is easily that of the [O III] 5007\AA\ line (fig \ref{SB_profiles}, top right), as expected \citep{Rappaport94}. 

In order to provide reference values for the expected characteristics of SSS nebulae, in table \ref{lines} we provide the total luminosities in the strongest optical emission lines (as well as the important diagnostic line He II 4686\AA), predicted for varying luminosities and temperatures of a central blackbody source, varying also the surrounding ISM density. Also shown for each combination of parameters in table \ref{lines} are the radial distances at which Hydrogen is 50\% ionized ($\rm{R}_{HII,0.5}$), and at which 50\% of Helium is in the singly-ionized state ($\rm{R}_{HeII,0.5}$). These results include input source parameters equivalent to models 1 through 5 of \cite{Rappaport94}, from which it is clear that the output from Cloudy is consistent with their work (within a few \% for ($\rm{R}_{HII,0.5}$) and ($\rm{R}_{HeII,0.5}$), $\lesssim$30\% for all emission lines). However, table \ref{lines} also extends to much lower densities then previously considered, and may be much more easily interpolated for comparison with observations. 

Inspecting our results, we see that the total line fluxes are relatively insensitive to the density of the surrounding ISM, with both the He II 4686\AA\ and H$\alpha$ emission decreasing only modestly as n falls from 10$\rm{cm}^{-3}$ to 0.1$\rm{cm}^{-3}$. At the same time, a drop by two orders of magnitude in the ambient ISM density results in a $\sim$50\% boost to the predicted [O I] 6300\AA\ and [N II] 6548\AA+6584\AA\ doublet emission, while the total luminosity reprocessed into the [O III] 5007\AA\ line is halved. Nonetheless, even in the lowest density case the [O III] 5007\AA\ forbidden line remains the most prominent single line for lower SSS temperatures ($\approx 2\times 10^{5}K$), lending further credence to focusing on this line in observational searches for SSS nebulae. Higher temperature ($\approx 10^{6}K$) sources may more readily reveal themselves in [O I] 6300\AA\ emission, although such objects should also be detectable in X-rays in the absence of significant obscuration. Altogether, it is evident from table \ref{lines} that the principle impact of lower ISM densities is the much more extended nebular radii predicted, and the correspondingly much lower surface brightnesses.

\subsection{Towards a more realistic model}

Past efforts to search for nebulae surrounding SSSs have found only one, that of CAL 83 \citep{PM89, Remillard95, Gruyters12}. The inferred density for the inner region of this nebula is $\approx 4$--$10\rm{cm}^{-3}$. For the 9 other Magellanic SSSs then known, \cite{Remillard95} report an upper limit of $\lesssim 10^{34.6}$erg/s on the [O III] 5007\AA\ luminosity for any associated nebula. Note however, that this upper limit was obtained specifically for an aperture with radius 7.5 pc, see the following section. In the context of the simple model presented above, this suggests that either the ambient gas density is much lower in the vicinity of most sources (recall eq. \ref{lines}, and note the typical radii for low ISM densities in table \ref{lines}), or that perhaps most ``SSSs'' spend the majority of their accretion histories at much lower effective temperatures, implying some failure of our understanding of the radial response of WD photospheres to accretion. Indeed, many SSSs are observed to be variable to differing degrees \citep[][]{Greiner2000}.  In this case, the detection of surrounding emission lines may trace the time-averaged behaviour of the source, revealing the past accretion history \citep{CR96}. Such variability is likely to explain in part the difficulty in fitting the CAL 83 nebula with any single-temperature model \citep{Gruyters12}. 

However, the simple picture presented above does not make mention of the density structure of the warm-neutral/warm-ionized ISM in star-forming galaxies. Furthermore, the photoionization calculations of \cite{Rappaport94} and \cite{Remillard95} make the standard assumption of an initially neutral medium. This is reasonable for classical H II regions powered by O stars, well-shielded as they are from any ambient radiation field by the dense clouds in which they are born. However, since SSSs are a much older population \citep[with ages likely peaking around 1Gyr, e.g.][]{Chen14}, there is no reason to assume they should typically be associated with dense H I clouds, but are rather immersed in the diffuse intercloud medium.

A detailed treatment of the resulting nebulae requires modeling the 3D distribution of interstellar matter and ionizing sources, which we reserve for a future study. In the following section, we investigate the structure of the ISM in the Magellanic clouds, focussing on the LMC and its 6 close-binary SSSs \citep{Greiner2000}, in order to estimate the typical densities in the vicinity of SSSs, and compare this with the assumption made in \cite{Remillard95}.

\begin{table*}
\renewcommand{\arraystretch}{1.1}
\begin{tabular}{c c c c c c c c}
\\
\hline 
Label & Name & $\rm{L}_{\rm{bol}}$ & $\rm{T}_{\rm{eff}}$ &  $\rm{N}_{\rm{H}}^{\rm{X}}$  & $\rm{N}_{\rm{H}}^{\rm{MW}}$ & ${N}_{\rm{H}}^{\rm{MC,tot}}$ & $n_{0}$\\
& & $10^{37}$ erg/s & eV & $10^{21}\rm{cm}^{-2}$ &$10^{21}\rm{cm}^{-2}$ & $10^{21}\rm{cm}^{-2}$ & $\rm{cm}^{-3}$\\
(1) & (2) & (3) & (4) & (5) & (6) & (7) & (8)\\
\hline
{\bf LMC} & & & & & & & \\
\hline
A & RXJ0439.8-6809	& 10-14 & 20-30 & 0.25 -- 0.4 & 0.45 & 0.28 & 0.22\\
B & RXJ0513.9-6951	& 0.1-6 & 30-40 & 0.4 -- 0.9 &  0.84 & 0.70 & 0.55\\
C & RXJ0527.8-6954	& 1-10	& 18-45 & 0.7 -- 1.0	 & 0.62 & 0.78 & 0.61\\
D & RXJ0550.0-7151	& - & 25-40 & 2.0  & 0.89 & 1.04 & 0.81\\
E & CAL 83		&  $<$2	& 39-60 & 0.7 -- 0.95 &  0.65  & 2.02 & 1.57\\
F & CAL 87		& 6-20	& 50-84 & 0.7 -- 2.1	 & 0.75  & 2.90 & 2.26\\
\hline
{\bf SMC}  & & & & & & & \\
\hline
G & 1E 0035.4-7230	& 0.5--4 & 30 -- 50 & 0.2-0.8 & 0.65 &  0.30 & -\\
H & RXJ0058.6-7146	& - & 15-70 & 0.3-1.5 & 0.75 & 3.14 & -\\
\hline
\end{tabular}
\caption[Known SSSs in the Magellanic Clouds]{Known SSSs in the Magellanic Clouds. Column 1 identifies each source with the corresponding label/location in fig. \ref{HI_MCs}. Cols. 2 -- 6 give the source name and measured values as tabulated by \protect\cite{Greiner2000}: col 3 gives the estimate of the bolometric luminosity, col 4 the best-fit temperature, col 5 the H I column density from fitting the X-ray spectra, and col. 6 the galactic column density in the direction of the source. Note that no luminosities are given for those sources reported as being poorly fit or constantly declining in the Greiner catalog. Col. 7 provides the total H I column density associated with the MCs, as measured by integrating all H I 21cm emission \protect\citep[from][]{GASS09, GASS10} with velocities $>$100 km/s along the line of sight of each source. Column 8 gives the inferred density in the midplane, assuming the H I in col 7 is distributed in an isothermal disk as discussed in the text. No value for $n_{0}$ is given for the two SMC sources, as the SMC is not a disk galaxy. \label{SSSs}}
\end{table*}

\section{SSS nebulae in the Magellanic clouds}\label{SSSs_MCs}

\subsection{A qualitative summary of ISM structure in the Magellanic clouds}

\begin{figure}
\begin{center}
\includegraphics[height=0.25\textheight]{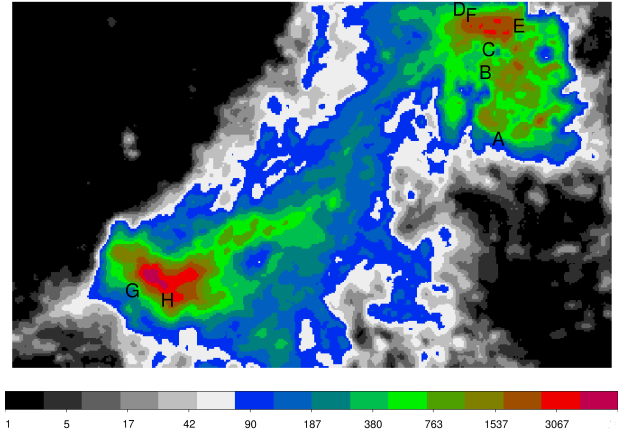}
\caption[H I 21cm map of the Magellanic clouds]{H I 21cm map of the Magellanic System \citep{GASS09, GASS10}. Units are in K$\frac{\rm{km}}{\rm{s}}$.  In the optically thin case (valid for almost the entirety of the Magellanic Clouds) this gives the total column density with the simple scaling 1.82$\times 10^{18}\rm{cm}^{-2}(\rm{K})^{-1}(\rm{km/s})^{-1}$. Labels denote locations of known SSSs, excluding PNe and symbiotics; letters correspond to sources in table \ref{SSSs}.}\label{HI_MCs}
\end{center}
\end{figure}

H$\alpha$ \citep{Kennicutt95} and H I 21cm maps \citep[][see fig. \ref{HI_MCs}]{GASS09, GASS10} of the Magellanic clouds reveal a complex, multiphase and multiscale ISM. Column densities in the LMC range from $\approx 4.5 \times 10^{21}\rm{cm}^{-2}$ near the centre to $\approx 10^{20}\rm{cm}^{-2}$ in the outer extremities \citep{Bruens05}. The distribution of the neutral ISM with scale height above the disk may be estimated from the velocity dispersion of H I, which \cite{Kim99} give as $\sigma$ $\approx$ 7.3 km/s. Assuming an isothermal disk for the LMC, they then find a scale height of $z_{0}$ $\approx$ 0.18 kpc. From this, we can make an estimate of the {\it mean} density of the neutral ISM along any line of sight $l$ from the relation:

\begin{equation}
\rm{N}_{\rm{H I}} = \int_{-\infty}^{\infty} n (l) dl = \frac{2}{\cos(i)}\int_{0}^{\infty} n_{0} \rm{sech}^{2}\left(\frac{z}{z_{0}}\right) dz
\end{equation}

\noindent where the factor $\cos(i)$ accounts for the fact that the disk is not face-on \citep[the LMC is in fact inclined at $i \approx 30^{\circ}$, e.g.][]{Nikolaev04}. From the column densities typical of the LMC, we find a range in peak densities (i.e. in the midplane, $n_{0}$) of $\sim$ 0.1--4$\rm{cm^{-3}}$. Note that from fig. \ref{HI_MCs}, one may then infer that, relative to CAL 83, in this simple picture SSSs are indeed likely to be in relatively low-density media, at least as best as can be ascertained from the 2D projection on the sky (see also table \ref{SSSs} for a quantitative comparison of column densities, and peak densities for a smooth, isothermal disk profile).

In practice however, this provides only a crude picture of the interstellar medium in the Magellanic clouds, and in star-forming galaxies in general. It assumes a smooth distribution, when in fact neutral gas in the Magellanic clouds displays a complex, fragmentary structure \citep[e.g.][]{Stanimirovic99}, with a significant fraction of H I being concentrated in dense cold clumps or filaments with a correspondingly low volume filling fraction \citep[$<$5\% in the Milky Way,][]{DL90}. The space between these filaments and clouds is filled with warm-ionized/warm-neutral ISM, with densities of order $\sim$ 0.1 $\rm{cm}^{-3}$, and hot supernova-heated ISM, with densities of order $\sim$ 0.01 $\rm{cm}^{-3}$ \citep{Cox05}. The rarefied hot interstellar medium occupies  a significant fraction of the volume of the disk \citep[$\approx$50\% in the Milky Way, e.g.][]{Cox05}, with warm-neutral/warm-ionized ISM occupying much of the remainder.
 From H I 21-cm emission alone, it is at best difficult to distinguish the contributions from warm and cold neutral gas. However, cold H I has been estimated to account for roughly 1/3 of the total H I mass in the LMC \citep{Dickey94}, with much of this concentrated in the central region of the LMC associated with 30 Doradus.

\subsection{Probability of the occurrence of a detectable SSS nebula -- qualitative consideration}

The much lower densities in the bulk of the volume  discussed above provide a natural explanation for the difficulty in observing nebulae around most SSSs. 

For the characteristic  inter-cloud density of $\sim 0.1$ cm$^{-3}$,  a typical accreting, nuclear-burning WD would have a Str{\" o}mgren radius of $\sim$150 pc ($\sim 10\arcmin $ at the distance of the LMC), and correspondingly very low surface brightness (fig. \ref{SB_profiles}). SSSs embedded in hot ISM would produce no nebulae at all in their immediate vicinity. 

\begin{figure}
\begin{center}
\includegraphics[height=0.25\textheight]{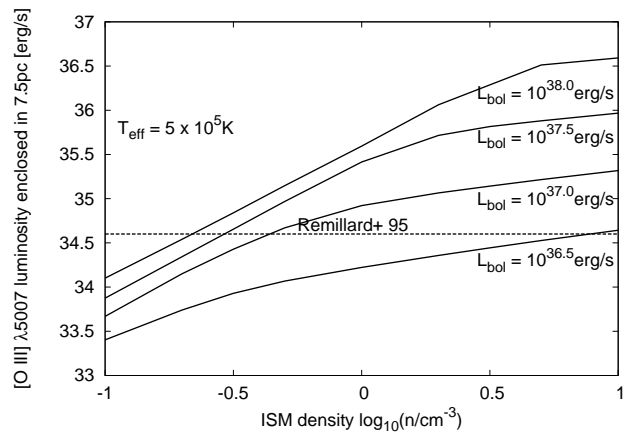}
\caption[Comparison with the results of Remillard et al, 1995]{Total [O III] 5007\AA\ luminosity enclosed within a 7.5pc radius as a function of (constant) surrounding ISM density. Different curves correspond to nebulae ionized by 5$\times 10^{5}$K blackbody sources with differing source luminosities. For comparison, the upper limit given by \protect\cite{Remillard95} is indicated by the dashed line.}\label{Check_Rem}
\end{center}
\end{figure}

\cite{Remillard95} do not detect [O III] 5007\AA\ nebular emission surrounding any of the Magellanic SSSs other than CAL 83. Their quoted upper limit on the luminosity in this line of $10^{34.6}$erg/s is given for an aperture radius of 7.5 pc, corresponding to a limiting surface brightness at the distance to the LMC of $\approx3\times 10^{-17}$erg/s/$\rm{cm}^{2}$/$\rm{arc sec}^{2}$. However, as can be seen from fig. \ref{SB_profiles}, this is not sufficient to exclude any nebula in the low-density ISM which must be typical of $\gtrsim$95\% of the LMC's volume. 
This is further illustrated in fig. \ref{Check_Rem}, where we plot the [O III] 5007\AA\ luminosity enclosed in an aperture with a radius of 7.5 pc as a function of ISM density, computed using our Cloudy models as above for a ``typical'' SSS (T = $5\times 10^{5}$K) with differing luminosities ($\rm{L}_{\rm{bol}}$ between $10^{36.5}$ and $10^{38}$ erg/s). For comparison, we plot the upper limit given by \cite{Remillard95}. As one can see from the plot, for the spherically symmetric case, SSS bolometric luminosities $\rm{L}_{\rm{bol}} \lesssim 10^{37.5}$erg/s remain consistent with \cite{Remillard95} upper limits  for $\rm{n}_{\rm{ISM}} \lesssim 0.4\rm{~cm}^{-3}$. 

Furthermore,  scale height of the intermediate and old stellar populations \citep[as traced by the radial velocity dispersions of carbon stars and old variable stars, respectively,][]{vdMarel06} are larger than that of the neutral ISM (and roughly consistent with the warm ionized medium). If the peak delay-time for the production of SSSs is $\sim$ 1 Gyr \citep{Chen14}, then many are likely to lie off of the midplane, in the lowest density neutral and ionized interstellar medium. Given the very large Str{\" o}mgren radii implied by such densities, a considerable fraction of ionizing photons originating from SSSs may then escape the galaxy entirely. Estimates of the surface brightness, and the corresponding constraint on the [O III] 5007\AA\ luminosity of known sources would then change accordingly.

The nebula associated with the SSS CAL 83, which has an inferred nebular density of $\approx$4--10$\rm{cm}^{-3}$ \citep{Remillard95, Gruyters12}, is overdense compared with that expected in the bulk ($\gtrsim 90\%$)  of the volume of the LMC. Given that the total mass ionized by the central source is estimated at roughly 150$\rm{M}_{\odot}$ \citep{Remillard95}, this cannot simply be the result of an outflow from the inner binary. The natural conclusion, as discussed above, is that the unusually compact nebula surrounding CAL 83 may be the product of a chance encounter  with a dense cloud or filament of cold neutral medium (hereafter CNM).  This raises the question of how likely such an encounter would be. Estimates for the volume filling fraction of the CNM vary from $\approx$1--5\% in spiral galaxies like the Milky Way \citep{DL90, Cox05}. For six known SSSs in the LMC, simple application of the binomial distribution would then suggest a probability of $\approx$ 97--99\% for $\leq$1 SSS lying within the CNM
($\approx$6--23\% that precisely one SSS lies inside a cold neutral cloud), assuming SSSs and the CNM are distributed equally throughout the LMC.

\subsection{A quantitative estimate of the proximity of cold neutral clouds}

The analysis presented above does not take into account the  distribution of cloud sizes, and only considers the likelihood that a SSS may lie {\em inside} a cold cloud. This does not account for the presence of dense ($\rm{n}_{\rm{CNM}}\sim 30$--$50\rm{cm}^{-3}$) cold clumps of H I which may typically lie sufficiently {\em near} to SSSs so as to be ionized and produce an associated nebula. If the number density of small cold clouds is sufficiently high, then many SSSs may be near enough to high-density regions to produce a detectable nebula, rendering the simple probability calculation presented above inaccurate.  In practice, what we require then is the distribution of distances to the nearest cloud surface for a SSS located at any arbitrary position within the volume of a galaxy. In the following, we will derive this employing a probabilistic argument with a number of simplifying assumptions; a more realistic treatment will require a detailed simulation of the multiphase ISM populated with SSSs, which we leave for a future study. In particular, we will treat all CNM clouds as spherical condensations in line with the two-phase model of \cite{FGH69}. In practice, the observed H I gas in the Magellanic clouds (and in our Galaxy) is now known to consist of many filaments, shells, and sheet-like structures, as well as clouds \citep{Cox05, Kim07}; however, this heuristic approach will allow us to make a first approximation.

We begin by considering the distribution of H I cloud sizes in the Magellanic system.  Summing the total area of any contiguous H I structure connected in RA, DEC, and velocity in the LMC, and identifying the square root of this area as the appropriate length scale, \cite{Kim07} find a distribution of H I cloud sizes which follows a roughly power-law distribution:

\begin{equation}
dN = Cl^{-\rm{\gamma}}dl,\hskip3cm 2.2 < \gamma < 3.0 \label{sizes}
\end{equation}

\noindent from kpc scales down to the minimum resolution of their survey ($\approx$ 15pc), and with larger values of $\gamma$ for lower flux-limits. Similar results are found for the SMC \citep{Stanimirovic99}, and observations in our own Galaxy as well as theoretical models of the formation and destruction of cold H I clouds suggest this distribution extends down to subparsec sizes \citep[][and references therein]{MO77, DL90}. In order to solve for the coefficient C, we assume spherical clouds and integrate over the volume of all spheres (for some reasonable choice of maximum and minimum cloud radii), and normalize this to the volume filling fraction of the CNM (per $\rm{pc}^{3}$): 

\begin{equation}
C = \frac{3(4-\gamma)}{4\pi}\frac{f_{\rm{fill}}}{l_{\rm{max}}^{4-\gamma} - l_{\rm{min}}^{4 - \gamma}}\label{coef}
\end{equation}

\noindent As discussed above, H I 21cm surveys alone cannot distinguish between the warm and cold phases of the neutral ISM, therefore both are included in the distribution given in eq. \ref{sizes}. However, both in our Galaxy and the Magellanic clouds, the total mass in H I clouds is found to vary with their size as $\sim l^{2}$ \citep[albeit with considerable scatter,][]{DL90,Kim07}, suggesting that the characteristic gas density in clouds of size $l$ should scale roughly as $\sim l^{-1}$. Observations in the solar neighbourhood suggest clouds of size $\sim$0.5pc have densities $\approx 30-50\rm{cm}^{-3}$, therefore one may infer an upper bound of l$\approx$50pc (and typical corresponding densities $\approx 0.3-0.5\rm{cm}^{-3}$) for the transition from CNM to warm-neutral medium length scales. We set the lower bound for cloud radii at 0.4pc, the theoretical lower limit on the size of small clouds found by \cite{MO77}. Note however that very small ($\sim$10--100 AU) clouds have also been detected in the Milky Way along many sightlines, which may comprise up to 15\% of the total H I column density in our galaxy \citep{Frail94}. However, these structures are thought to be physically unrelated to the ``standard'' CNM, arising instead from shocks surrounding individual low-mass stars and other local phenomena \citep{Stanimirovic99}. 

We can normalize eq. \ref{sizes} to the total number in order to obtain a probability distribution:

\begin{equation}
p_{\rm{l}}(l) = (\gamma - 1)\frac{l^{-\gamma}}{l_{\rm{max}}^{1-\gamma} - l_{\rm{min}}^{1-\gamma}} 
\label{prob_sizes}
\end{equation}

We would then like to find the distribution of distances from an arbitrary point to the nearest cloud surface.  At first glance, this can be straightforwardly computed via convolution of eq. \ref{prob_sizes} with the distribution of distances to the nearest cloud centre:

\begin{equation}
p_{D}(D) = \int_{l_{\rm{min}}}^{l_{\rm{max}}} p_{\rm{r}}(D + l)p_{\rm{l}}(l)\theta(D + l)dl \label{simple}
\end{equation}

\noindent with $D = r-l$ the distance to the nearest cloud surface and $p_{\rm{r}}$ being the standard inter-particle spacing distribution $p_{r}(r) = 4\pi r^{2}n_{\rm{clouds}}\rm{e}^{-\frac{4\pi}{3}n_{\rm{clouds}}r^{3}}$ \citep{Chandra43}. However, this simple treatment 
is only accurate in the limit $f_{\rm{fill}} \rightarrow 0$, as it does not account for the overlapping of spheres \citep[see e.g.][for a discussion of the analogous problem in the study of concrete porosity]{Snyder98}.  In practice, it  fails for values of the volume filling factor as small  as a $\sim$few per cent\footnote{The reason  is that two spheres overlap  when the distance between their centres becomes smaller than $r_1+r_2$. Therefore the effective filling factor in estimating the importance of overlapping is $2^3\times f_{\rm fill}$, which equals 0.4 for $f_{\rm fill}=0.05$.}. For example, at $f_{\rm{fill}}$= 0.05, integration of eq. \ref{simple} gives a cumulative probability for D$<$0 (i.e. of a random point lying within a cloud) of only $\approx$1\%, although the correct value is obviously equal to $f_{\rm{fill}}$= 5\%.

\begin{figure}
\begin{center}
\includegraphics[height=0.25\textheight]{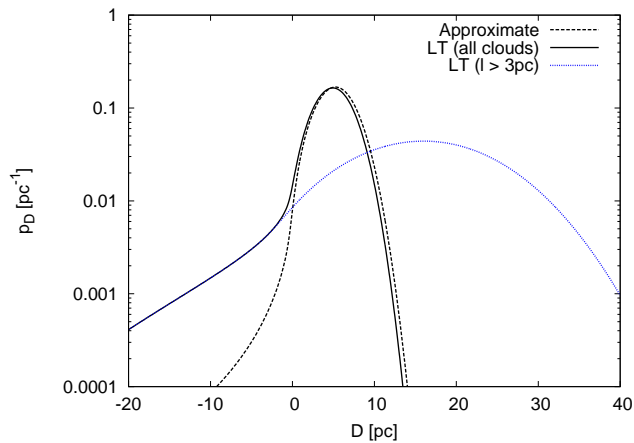}
\caption[Nearest surface probability density distribution]{Nearest surface probability density distribution ($p_{\rm{D}}$) for CNM clouds in the ISM. The three curves correspond to $p_{\rm{D}}$ as estimated from eq. \ref{simple} (dashed line), and using the ``hard spheres'' formulation from \protect\cite{LT92} (solid line). Also shown is the nearest surface probability density distribution for only those CNM clouds with l $>$ $3$pc (blue dotted line). In all cases, we have assumed a {\it total} volume fraction $f_{\rm{fill}}$ = 0.05, with $\gamma$ = 3, $l_{\rm{min}}$ = 0.4pc and $l_{\rm{max}}$ = 50pc.}\label{Prob_D}
\end{center}
\end{figure}

In order to improve on this, one must modify equation \ref{sizes} in order to enforce the separation of clouds. This is a well-studied problem which arises often in materials science and chemistry; a more precise formulation is that of \cite{LTR90} (for mono-sized inclusions) and \cite{LT92} (for polydisperse clouds); its main results are summarized in Appendix A. The correct differential distribution of distances to the nearest cloud surface from  Appendix A is plotted in fig. \ref{Prob_D} for $f_{\rm{fill}}$=0.05 and $l_{\rm{min}}$ and $l_{\rm{max}}$ as above. From eq. \ref{LT}, we find a probability of $\approx$86\% that a CNM cloud surface lies within 7.5pc of any SSS \citep[and thus within the aperture of][]{Remillard95}. This is sensible: with a number density of $\approx4\times10^{-4}$ cloud centres per cubic parsec, a spherical volume with radius 7.5pc should contain $\sim$1 cloud centre. However, from eq. \ref{sizes} above, it is clear that the overwhelming majority of such clouds should be very small ($<$1pc). In order to estimate the likelihood of a nebula such as that detected around CAL 83, we must then ask: what is the likelihood of a SSS being in close proximity to the surface of a cloud which is large enough to encompass a total mass of hydrogen which will be detectable when ionized? 

For spherical clouds, the radius required to enclose a given mass for fixed density is simply:

\begin{equation}
l = 0.69\rm{pc}\left(\frac{n}{30\rm{cm}^{-3}}\right)^{-\frac{1}{3}}\left(\frac{M}{M_{\odot}}\right)^{\frac{1}{3}}
\end{equation}

\noindent The total mass of the nebula surrounding CAL 83 has been estimated at $\approx$150$M_{\odot}$ \citep{Remillard95}. A CNM cloud with an initial density of 30$\rm{cm}^{-3}$ (i.e. typical of the neutral gas, prior to being ionized) would need to be $\approx$3pc in size in order to supply the requisite material. Therefore, in fig. \ref{Prob_D} we also plot the probability density function for the nearest cloud surface for clouds with radii $>$3pc. Once ionized, the expansion timescale of the formerly cold cloud is roughly the length over the sound speed in the ionized cloud, or 

\begin{equation}
\tau _{\rm{expansion}} = \frac{l}{c_{\rm{i}}} \approx 10^{6} \left(\frac{l}{10\rm{pc}}\right) \left(\frac{c_i}{10{\rm~km/s}}\right)^{-1} \rm{years}
\end{equation}
This is comparable to the lifetime of the SSS itself, so it is unsurprising if the observed ionized nebula is not in hydrostatic equilibrium with the surrounding warm ISM \citep[see also][]{CR96}.

These larger clouds occupy much of the volume (cf. eq, \ref{sizes}), therefore the cumulative probability of landing within a cloud (D $<$ 0) with l $<$ 3pc is only slightly reduced (P(D$\leq$0) $\approx$ 0.0473). The probability of such a CNM cloud lying within 7.5pc is $\approx$18\%. We may then conclude that out of six SSSs, only one having a detectable associated nebula appears entirely consistent with our crude model of the CNM; from the binomial distribution, the likelihood of exactly 1 out of 6 known SSSs lying near a cloud is 40\%.  Therefore, on the basis of (presently very limited) published observations, one cannot make any statement of the long term time-averaged luminosities and temperatures of known SSSs without detected ionized nebulae.

\section{Prospects for detecting SSS nebulae}\label{Prospects}

\subsection{Searching for previously undetected nebulae ionized by SSSs}

The question then becomes whether deeper narrow-band observations could detect nebulae in lower density ISM, as well as possibly reveal a SSS population, including perhaps many accreting, nuclear-burning WDs, unseen in soft X-rays (see $\S$\ref{Implications}). For detecting relatively low-temperature ($\sim 2\times10^{5}$K) sources, the ideal line to focus on may be [O III] 5007\AA, as this is predicted to be the strongest optical line surrounding these objects (recall $\S$2.2). One starting place would be to re-examine past planetary nebulae surveys, which would have rejected any extended objects. This could be complimented by a renewed search using a large aperture telescope such as Magellan or the VLT, particularly in the vicinity of known SSSs. Narrow-band searches centred on [O I] 6300\AA\ would be necessary for the discovery of nebulae ionized by low-luminosity, higher temperature sources ($L \sim 10^{37}$erg/s, $T\sim 10^{6}$K). Follow-up spectroscopy of any newly discovered, faint and extended objects would be able to shed light on the nature and number of such sources.

To determine the feasibility of this, we can estimate the exposure time needed to detect a nebula of constant density n surrounding an ionizing source of some fixed temperature and luminosity. For a line flux $\rm{F}_{\rm{S}}$ which, for a given nebula, is spread over some solid angle with surface brightness $I_{S}$, the measured ``signal'' at a telescope on the Earth's surface (in electrons per second per pixel) is \citep[see e.g.][]{AstroLecture}:

\begin{equation}
C_{\rm{S}} \approx \frac{\pi}{4}\left(\frac{D_{\rm{aperture}}}{4.86\times
10^{-6}L}\right)^{2}\frac{\rm{I}_{\rm{S}}}{h\nu}\langle\rm{T}\rangle
d_{\rm{pix}}^{2}
\label{measured_counts}
\end{equation}

\noindent for a given telescope with an aperture diameter $\rm{D}_{\rm{aperture}}$, focal length L, physical pixel size $d_{\rm{pix}}$, line energy $h\nu$, and effective filter transmission $\langle T \rangle$. Note that we make the simplifying assumption here of a single effective mean transmission, independent of wavelength, and a single line energy. 

In an observation this flux is integrated over an exposure time $t$ and a total number of source pixels $N_{\rm{pix,S}}$. The expected S/N of such a measurement may be estimated given some background flux (to be subtracted from the measured flux), as well as the read noise $RN$ and the dark current $DN$, which characterize contributions to the uncertainty from the CCD detector itself. We adapt the standard formula \citep[see e.g.][]{CCDtext}:

\begin{equation}
\frac{\rm{S}}{\rm{N}} =
\frac{\sqrt[]{N_{\rm{pix,S}}} C_{\rm{S}} t}{\sqrt{C_{\rm{S}} t + (1 + \frac{N_{\rm{pix,bkg}}}{N_{\rm{pix,S}}})(C_{\rm{bkg}}\langle\rm{T}\rangle t + (DN)t + N_{\rm{DIT}}(\rm{RN})^{2})}}\label{SN_eq}
\end{equation}

\noindent with $N_{\rm{DIT}}$ the total number of exposures (1 unless otherwise indicated), $N_{\rm{pix,bkg}}$ the number of pixels over which the background is measured, and $C_{\rm{bkg}}$ the background signal assuming some mean surface brightness $I_{\rm{bkg}}$ (where $C_{\rm{bkg}}$ is computed from $I_{\rm{bkg}}$ in a matter analogous to eq. \ref{measured_counts}). For the present estimates, we fix the ratio $\frac{N_{\rm{pix,bkg}}}{N_{\rm{pix,S}}} = 3$, although this may prove difficult for nebulae with the largest radii. Note that, in order to avoid the inclusion of any faint nebular emission in evaluating the background, it would be best to estimate the background far from the source, although care must be taken in selecting an appropriate region. 

For the purposes of evaluating the detectability of SSS nebulae given modern capabilities, we shall assume narrow-band observations carried out with the Magellan Baade Telescope.\footnote{For reference, see http://www.lco.cl/telescopes-information/magellan/} We focus on the [O III] 5007\AA\ line, as this is the strongest optical emission line in our calculations. We assume an effective filter transmission of $\langle T\rangle$ $\approx 80$\%, with $RN$ $\approx$ 4.1$\rm{e}^{-}$, and $DN$ = 5$\rm{e}^{-}$ per pixel per hour (Alejandro Clocchiatti, private communication). In order to estimate the surface brightness of a nebula, we compute the [O III] luminosities and nebular radii for varying SSS luminosities and ISM densities, assuming a photospheric temperature of T = $2\times 10^{5}$K. Note that we take a constant mean surface brightness as representative of the nebular emission, with the outer radius defined by the point at which the gas temperature falls below 3000K (recall $\S$\ref{Model_SSS_Nebulae}). This will underestimate the surface brightness in the bulk of most nebulae, providing a conservative estimate of their detectability. The characteristic surface brightness of the background flux $\rm{I}_{\rm{bkg}}$ is less certain, and is not generally well-approximated by a constant value in a star-forming galaxy. For our theoretical estimates presented here, we assume $\rm{I}_{\rm{bkg}}=5\times 10^{-16}$erg/s/$\rm{cm}^{2}$/$\rm{arcsec}^{2}$, which \cite{Pellegrini12} found to be a representative mean in their H$\alpha$ images of the Magellanic Clouds. As the optical continua of irregular galaxies are fairly flat \citep{Kennicutt92} we take this as our estimate for the background continuum for all lines of interest discussed in $\S$\ref{SSSs_MCs}. We ignore the sky background, as this is generally minimal for optical narrow-band observations. Note that, for a real measurement, consideration of the largest possible Str{\" o}mgren radii may introduce difficulties in finding a suitable region to measure the background in a single observation; however, the problem may be solved with multiple observations covering a larger area. 

\begin{figure}
\begin{center}
\includegraphics[height=0.25\textheight]{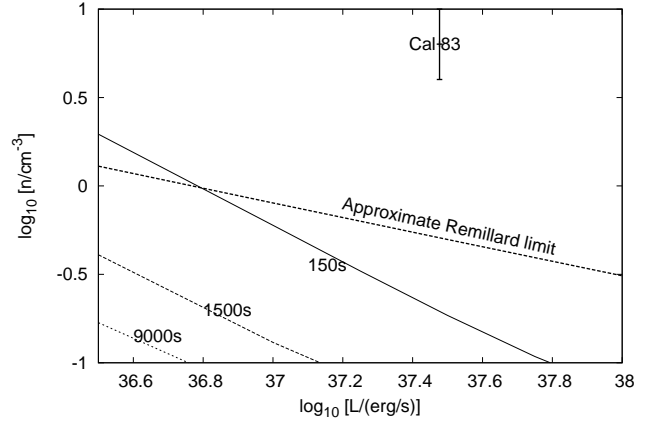}
\caption[Detectability of SSS nebula in the LMC]{ISM density required to produce a detectable nebula ionizing by a $2\times 10^{5}$K source with bolometric luminosity L. The three curves denote a signal to noise ratio of 50 for 150, 1500, and 9000 seconds total integration times using Magellan, respectively. For the latter, we assume the observation is carried out in 6 exposures (cf. eq \ref{SN_eq}). For reference, the inferred density and time-averaged luminosity of CAL 83 is also shown, as is the surface brightness limit inferred for the upper limits on the [O III] 5007\AA\ line luminosity given in \protect\cite{Remillard95}. } \label{SN}
\end{center}
\end{figure}

For a given source luminosity and temperature, we may then ask what surrounding ISM density would be required in order to produce a sufficiently compact and bright nebula, such that it may be detected at a given S/N in a reasonable exposure time. 
Given that theoretical S/N estimates present only the best case scenario (with all sources of error accounted for), here we define a "confident detection" as a measurement with S/N = 50, given a signal integrated over the entire nebula for any combination of parameters.  Plotted in fig. \ref{SN} are the minimum surrounding ISM densities required for an [O III] nebula to be detected around a T = $2\times 10^{5}$K SSS of a given luminosity, for differing exposure times. Also shown for reference is the approximate limiting surface brightness inferred for the [O III] 5007\AA\ and H$\alpha$ upper limits given in \cite{Remillard95}. Note that the slope of the latter is somewhat steeper than may be expected from eq. \ref{RS} and with a fixed limiting surface brightness. This is because the total luminosity in [O III] 5007\AA\ does not depend exclusively on the bolometric luminosity, but also (weakly) on the ambient ISM density.  We see that given modern capabilities \cite[i.e. the 6.5m Magellan telescopes, as opposed to the 1.5m aperture available to][]{Remillard95}, one could in a $\approx$ 1500s exposure detect any SSS nebula with $L_{\rm{bol}} \gtrsim 10^{37.1}$ erg/s down to ambient ISM densities $n \gtrsim 0.1\rm{cm}^{-3}$. 

\subsection{Identifying ``orphan'' high-ionization regions as possible SSS nebulae}

One may also investigate known nebulae without identified ionizing sources which appear to have emission-line ratios and line fluxes consistent with ionization by a SSS. Although difficulties remain in modeling the observed optical emission-line ratios in the CAL 83 nebula, the basic properties (He II flux, its ratio to H$\beta$, and radial extent of the nebula, given the inferred density) appear consistent with ionization by a SSS near its centre. From an intuitive standpoint, and informed by our estimates above, it is unlikely this is the only object of its kind. Indeed, two extended He II nebulae have recently been identified in M33 \citep{Kehrig11} which have no observed optical or X-ray counterpart (a Wolf-Rayet star, high-mass X-ray binary, etc.).  One of these, HBW 673, appears to have a He II Str{\" o}mgren radius of $\sim$ 20pc, with He II 4686\AA/H$\beta$ = 0.55 and a He II 4686\AA\ line luminosity of $\gtrsim 10^{33}$erg/s. This is consistent with ionization by a $\approx 10^{37\rm{-}37.5}$erg/s source with effective temperature 2--3$\times 10^{5}$K, embedded in a surrounding ISM with density $\approx 1\rm{cm}^{-3}$. Such a soft source could easily have evaded prior detection (Woods et al, in prep). 

The other nebula, BCLMP 651, has a lower He II 4686\AA/H$\beta$ ratio of 0.11; if the ionizing source is indeed an accreting, nuclear-burning WD, then the photospheric temperature must either lie below $\approx 1.5\times10^{5}$K or above $\approx 6\times 10^{5}$K \citep{Rappaport94, WG13}. The very low value of [O I]6300\AA/H$\alpha$ ($\approx 0.01$) and non-detection of any X-ray source favour the low-temperature case, however detailed photoionization modeling (and possibly X-ray and optical follow-up) are needed in order to resolve the nature of the sources ionizing both nebulae.

\section{Implications for the ISM of star-forming galaxies}\label{Implications}

If there exists a large population of previously unaccounted-for high-temperature ionizing sources, this may carry important consequences for our understanding of the ISM. Given the relative ubiquity of WDs, and the number of suspected WDs among known SSSs \citep[e.g.][]{Greiner2000}, one would expect the majority of these sources to be accreting, nuclear-burning WDs. Indeed, the SD scenario for the progenitors of SNe Ia suggests a large number of such objects, and present population synthesis calculations suggest the SD scenario may account for a significant fraction of the SN Ia rate observed at early delay-times, and in galaxies with vigorous ongoing star-formation \citep[in the latter, perhaps $\sim 10\%$ at $\sim 1$ Gyr, see e.g. ][]{Chen14}. In particular, the SD scenario predicts a minimum average total luminosity of the accreting, nuclear-burning WD population \citep{GB10}:

\begin{equation}
\rm{L}_{\rm{Ia}} \approx 9\times 10^{30} \left(\frac{\Delta \rm{M}}{0.3\rm{M}_{\odot}}\right)\left(\frac{\dot N_{\rm{Ia}}}{10^{-13}\rm{M}^{-1}_{\odot}\rm{yr}^{-1}}\right) \frac{\rm{erg}}{M_{\odot}\rm{s}}\label{SD_SNeIa}
\end{equation}

\noindent neglecting those accreting WDs which do not explode as SNe Ia. The SN Ia rate has been measured in the Magellanic clouds from observations of expanding supernova remnants \citep{MB10}, and found to be roughly consistent with (and perhaps much higher than) the rate in dwarf irregular galaxies, $\approx 8\times 10^{-13}$ SNe $\rm{yr}^{-1}\rm{M}_{\odot}^{-1}$ \citep{Mannucci05}. The total stellar mass of the LMC is $\approx 2.7\times 10^{9}\rm{M}_{\odot}$ \citep{vdMarel06}. Therefore, if SD progenitors accrete on average $0.3\rm{M}_{\odot}$ per WD at high photospheric temperatures ($\gtrsim 10^{5}$K), they would produce a total luminosity of 2$\times 10^{41}$ erg/s. This would principally be in the extreme-UV, with blackbodies in this temperature range ($2\times 10^{5}$K $<$ T $<$ $10^{6}$K) producing 3--12$\times 10^{9}$ ionizing photons per erg. We can make an estimate of the total H$\alpha$ luminosity this may produce (if no ionizing photons escape), assuming Case B recombination in the low-density limit, with a (typical) gas temperature of $10^{4}$K \citep{Osterbrock}. This gives us a predicted H$\alpha$ luminosity of 1--3$\times 10^{39}$erg/s, roughly 10--30\% of the total diffuse H$\alpha$ luminosity of the LMC \citep[i.e., excluding that immediately associated with resolved H II regions,][]{Kennicutt95}. This fraction of the total non-H II region H$\alpha$ emission is comparable to the ``truly diffuse'' H$\alpha$ luminosity found by \cite{Kennicutt95}, which is not associated with either filamentary structures, or the diffuse flux within supergiant shells. Therefore, understanding the long-term variability of SSSs may be a question of vital importance to the study of the structure of the ISM in star-forming galaxies.

\begin{figure}
\begin{center}
\includegraphics[height=0.25\textheight]{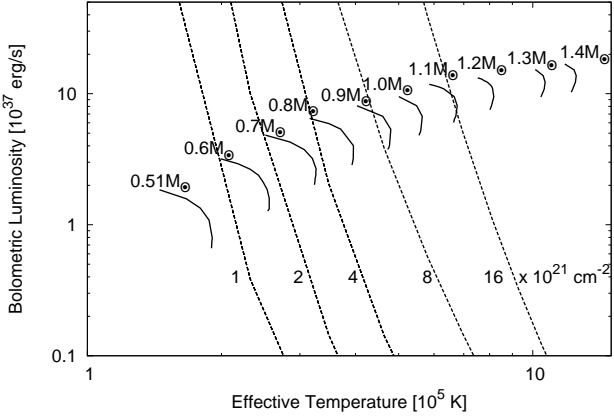}
\caption[Direct detectability of SSS nebulae]{Estimate of the minimum bolometric luminosity needed for an accreting, nuclear-burning WD of a given temperature to be detected as a SSS by XMM-Newton. Dashed lines correspond to different fixed column densities, consistent with values observed in the MCs (from left to right, $10^{21}$, $2\times 10^{21}$, $4\times 10^{21}$, $8\times 10^{21}$, and $1.6\times 10^{22}\rm{cm}^{-2}$. Also shown are the temperatures and luminosities of the steady-burning models computed by \protect\cite{WB13} (similar to that shown in their fig. 2) for accreting WDs of different masses (solid lines).
} \label{detecting_SSSs}
\end{center}
\end{figure}

The few observed SSSs in the LMC have a combined luminosity that is less than 1\% of the value suggested by eq. \ref{SD_SNeIa}, even if we (very optimistically) sum only their maximum luminosities. The question arises: is it possible that a large number of accreting, nuclear-burning WDs could remain undetected in the LMC? Shown in fig. \ref{detecting_SSSs} is an estimate of the minimum bolometric luminosity for accreting nuclear-burning WDs of a given temperature required for it to be detected in soft X-rays, assuming differing column densities consistent with that observed in the MCs. To compute this, we make a simple estimate assuming blackbody spectra and a minimum detectable count rate of $10^{-3}$ counts per second, consistent with the faintest confidently detected SSSs in the recent XMM-Newton point source catalog of the SMC \citep{Sturm13}. Note that, in using a single limiting threshold count rate, we are painting the MCs in one broad stroke.  For this reason these results must be considered to be accurate to within a factor of a few. Strictly speaking the current sensitivity at any point in the MCs would vary, and depends on the exposure times of all observations taken up to the present time. In order to compute the observed count rate, we made use of the command-line PIMMS tool\footnote{http://heasarc.gsfc.nasa.gov/docs/software/tools/pimms.html}, using the thin filter for the EPIC-PN instrument of XMM-Newton. Also shown are results from the 1D steady-burning WD models of \cite{WB13}.

Two things are apparent from inspection of fig. \ref{detecting_SSSs}: firstly, that it is clear all high-mass ($\gtrsim 1.1\rm{M}_{\odot}$) nuclear-burning WDs should now be detected in the MCs, unless during the majority of their accretion phase their photospheric radii deviate greatly from that expected in the 1D models of \cite{WB13}; second, that the population of low-mass steadily nuclear-burning WDs is, to a large extent, unconstrained by X-ray observations alone. This is apparent from comparison of the $N_{\rm{H}}$ contours in fig. \ref{detecting_SSSs} with the typical MC column densities in fig. \ref{HI_MCs}. Note that while most population synthesis models do not predict a dominant SD channel in the production of SNe Ia \citep{Nelemans13,Chen14}, they do predict a large number of low-mass accreting, nuclear-burning WDs nonetheless. In star-forming galaxies, recent population synthesis calculations indicate the total luminosity of {\it all} accreting, nuclear-burning WDs could provide a significant source of ionizing radiation, although it is substantially below the estimate given above for the luminosity of a putative SD progenitor population \citep{Chen14,Chen15}. By far the majority of accreting nuclear-burning WDs should lie at lower masses (0.6$\rm{M}_{\odot}$--$0.8\rm{M}_{\odot}$), most of which should still remain undetected in the MCs (see fig. \ref{detecting_SSSs}). Therefore, there may yet exist a considerable number of such low-mass, steadily nuclear-burning WDs which have not yet been detected. Searching for nebulae ionized by accreting WDs may be the only way to observationally investigate this further.

The above considerations, together with the relatively low ionization state of Helium in the ISM, suggest nuclear-burning WDs likely play a relatively minor role in the overall ionization balance of diffuse interstellar gas (the ``DIG''). 

However, even if SSSs are not a major component of the diffuse ionizing background in star-forming galaxies, they may play a role in the high [O III]/H$\beta$ ratios seen in the low-density (n$\lesssim$0.1$\rm{cm}^{-3}$) ISM \citep{Haffner09}. In particular, it appears that some heretofore unknown mechanism is needed to account for the additional heating ($\sim 10^{-26}$ erg/s/$\rm{cm}^{-3}$) inferred in the diffuse ISM well above the midplane \citep{Reynolds99}. Nuclear-burning WDs could naturally account for such a hardening of the ionizing continuum, should their numbers and distribution with scale height prove appropriate. This would require considerable further investigation before any definitive statement can be made. Given the high escape fraction, nuclear-burning WDs may also provide a component of the apparently ``missing'' contribution to the cosmic ionizing photon background from galaxies at low redshift \citep{Kollmeier14}, as well as contributing to the heating balance of the hot coronal gas (S. Cantalupo, private communication). In particular, ionization of He II (an important coolant in the hot ISM) by nuclear-burning WDs may keep the gas sufficiently hot so as to provide an additional component of stellar feedback. The feasibility of such an additional mechanism should be tested by inclusion in theoretical models \citep[such as][]{Cantalupo10, Kannan14}, coupled with deep observations of the ISM in the MCs (as outlined in $\S$\ref{Prospects}). 

\section{Conclusions}\label{Conclusions}

Two decades after the discovery of SSSs, the long-term behaviour of accreting, nuclear-burning WDs remains poorly understood. Past efforts to search for nebulae ionized by these objects detected only one such object, with the implication that either the SSS phase is extremely transient in real accreting WDs, or most SSSs lie in low-density (n$\lesssim 1\rm{cm}^{-3}$) media. This problem has lingered without resolution since the survey of \cite{Remillard95}.

While it remains possible that our understanding of the long-term behaviour of SSSs is incorrect, in this work we have demonstrated that most SSSs must lie in much lower density ISM than does CAL 83. In particular, for our simple model we showed that for the six known close-binary SSSs in the LMC, the likelihood that no more than one source lies sufficiently near to a dense cloud to produce a nebula which could have been previously detectable is $\approx$ 70\%. This probability has been computed based on the current understanding of the structure of interstellar medium in star forming galaxies. Therefore, given the current state of the observational evidence there is no reason to conclude that SSSs have in general a much lower duty cycle than presently inferred.

We then outlined the prospects for a future survey which may be able to reveal all bright SSSs in the Magellanic clouds. The outcome of such a study carries important consequences regardless of the outcome. If no enhancement in [O III] 5007\AA\ emission is seen in the vicinity of known SSSs, with limiting surface brightnesses of $\lesssim 10^{-17}$erg/s/$\rm{cm}^{2}$/$\rm{arcsec}^{2}$, then these objects are not well-described by our present understanding of accretion onto nuclear-burning WDs. Alternatively, the discovery of a large population of previously undetected ionizing sources, distributed throughout the warm-ionized ISM, may have profound implications for our understanding of the evolution of the ISM in star-forming galaxies. 

\section*{Acknowledgements}
The authors would like to thank the anonymous referee for their remarks, which improved the clarity and utility of this manuscript, J{\" u}rgen Kerp for kindly providing the H I map of the Magellanic clouds shown in fig. \ref{HI_MCs}, as well as Alejandro Clocchiati, Pierre Maggi, Saul Rappaport, and Armin Rest for their helpful discussions. MG acknowledges  partial support by Russian Scientific Foundation (RNF), project 14-22-00271.

\appendix

\section{Lu \& Torquato treatment of the distribution of nearest surfaces}

In order to find the probability distribution of distances to the nearest surface for an arbitrary system, Lu \& Torquato developed an approximation using n-point correlation functions and obtaining exact expansions for mono-sized spheres of arbitrary dimension \citep{LTR90}. This was then extended to the case of ``polydisperse'' sphere radii \citep{LT92}. For hard (impenetrable) spheres (i.e. with an interaction potential which is infinite within any given sphere and zero elsewhere), \cite{LT92} give the cumulative distribution function for distances to the nearest ``particle'' (cloud) surface $p_{D}(D) = -de_{V}(D)/dD$, where $e_{V}(D)$ is given by: 

\begin{equation}
e_{V}(D) = \left\{
  \begin{array}{lr}
  1 - \frac{4\pi}{3}n\langle (D + l)^{3}\theta(D + l)\rangle \hskip 2.05cm D < 0 \\ \\
  (1 - f_{\rm{fill}})\exp[-\pi n(aD + bD^{2} + cD^{3})] \hskip0.55cm D > 0
   \end{array}
   \right.\label{LT}
\end{equation}

\noindent where:

\begin{equation}
\langle (D + l)^{3}\theta(D + l)\rangle = \int_{l_{\rm{min}}}^{l_{\rm{max}}}(D + l)^{3}\theta(D + l)p_{l}dl
\end{equation}

\noindent and the coefficients a,b, and c defined as:

\begin{equation}
a = \frac{4\langle l^{2}\rangle}{1 - f_{\rm{fill}}}
\end{equation}

\begin{equation}
b = \frac{4\langle l\rangle}{1 - f_{\rm{fill}}} + \frac{12\xi_{2}}{(1 - \xi_{2})^{2}}\langle l^{2} \rangle
\end{equation}

\begin{equation}
c = \frac{4}{3(1-f_{\rm{fill}})} + \frac{8\xi_{2}}{(1 - f_{\rm{fill}})^{2}}\langle l \rangle
\end{equation}

\noindent with 

\begin{equation}
\xi_{k} = \frac{\pi}{3}n2^{k-1}\langle l^{k} \rangle
\end{equation}

\noindent where $\langle l^{k} \rangle$ is the expectation value of $l^{k}$:

\begin{equation}
\langle l^{k} \rangle = \int_{l_{\rm{min}}}^{l_{\rm{max}}} l^{k}p_{l}(l)dl
\end{equation}

\noindent with $p_{l}$ given by eqs. \ref{sizes} \& \ref{coef}.

\label{lastpage}

\end{document}